\documentclass[journal]{IEEEtran}
\usepackage{amssymb}
\usepackage{amsfonts}
\usepackage{booktabs}
\usepackage[pdftex]{graphicx}
\usepackage[cmex10]{amsmath}
\usepackage{algorithm}
\usepackage{algorithmic}
\usepackage{array}
\usepackage[caption=false,font=footnotesize]{subfig}
\usepackage{colortbl}

\begin{document}
\title{Group Based Interference Alignment}
\author{Yan-Jun~Ma,
        Jian-Dong~Li,~\IEEEmembership{Senior~Member,~IEEE}, Qin Liu, Rui~Chen
\thanks{This work was supported in part by the National Science Fund for Distinguished
Young Scholars under Grant 60725105, by the National Basic Research Program of China
 under Grant 2009CB320404, by the Program for Changjiang Scholars and Innovative
 Research Team in University under Grant IRT0852, by the Key Project of Chinese Ministry of Education under Grant
  107103, and by the 111 Project under Grant B08038.}
\thanks{The authors are with the State Key Laboratory of Integrated
Service Networks, Xidian University, Xi'an 710071, China
 (e-mail: \{yjm, jdli, qinliu\}@mail.xidian.edu.cn, rchenxidian@gmail.com).}}
\maketitle

\begin{abstract}
In the $K$-user single-input single-output (SISO) frequency-selective fading
interference channel, it is shown that the maximal achievable multiplexing gain
is almost surely $K/2$ by using interference alignment (IA). However, when the
signaling dimensions are limited, allocating all the resources to all users
simultaneously is not optimal. So, a group based interference alignment (GIA)
scheme is proposed, and it is formulated as an unbounded knapsack problem.
Optimal and greedy search algorithms are proposed to obtain group patterns.
Analysis and numerical results show that the GIA scheme can obtain a higher
multiplexing gain when the resources are limited.
\end{abstract}

\IEEEpeerreviewmaketitle

\section{Introduction}
\IEEEPARstart{I}{nterference} management is an important problem in wireless
system design. As an effective technique for interference management,
interference alignment (IA) is first considered in~\cite{ref1}, \cite{ref2}~as
a coding technique for the two-user multiple-input multiple-out (MIMO) X
channel. Using this scheme, it is shown that each user can obtain almost surely
a multiplexing gain (MG) of $1/2$ per channel use in the $K$-user SISO
interference channel (IC) \cite{ref3}. A beamforming matrices optimized IA
(called BF-IA) scheme is proposed in \cite{ref4} which can obtain a higher MG
than the IA scheme in \cite{ref3} at any given number of channel realizations.
IA scheme is also applied in cellular networks in \cite{ref5} which can boost
 system performance in some scenarios. However, when the signaling
dimensions are limited, allocating all the resources to all users
simultaneously is not optimal.

In this letter, a GIA scheme is proposed based on the BF-IA scheme which can
obtain a higher MG when the resources are limited.
\section{System Model and Preliminaries}
Consider the $K$-user frequency-selective fading IC model:
\begin{equation} \label{eq1}
\mathbf{Y}^{[k]}=\sum_{l=1}^K{\mathbf{H}^{[kl]}\mathbf{X}^{[l]}+\mathbf{Z}^{[k]}},~
\forall k\in \{1,\dots,K\}.
\end{equation}
$\mathbf{X}^{[l]}$ is the $M\times1$ input signal vector of the $l^{th}$
transmitter, and $\mathbf{Y}^{[k]}$ is the channel output at the $k^{th}$
receiver, where $M$ is the number of available frequency-selective channel
realizations. $\mathbf{H}^{[kl]}$ is the diagonal channel matrix between
transmitter $l$ and receiver $k$. We assume that all $\mathbf{H}^{[kl]}$'s are
known in advance at all the transmitters and all the receivers, and assume the
channel is time-invariant. $\mathbf{Z}^{[k]}$ is $M\times1$ additive white
Gaussian noise (AWGN) vector at the $k^{th}$ receiver, where all noise terms
are independent identically distributed (i.i.d.) zero-mean complex Gaussian
with unit variance. The definition of achievable sum-rate follows from
\cite{ref3}. Define
$r=\lim_{\textrm{SNR}\to\infty}{\frac{R(\textrm{SNR})}{\log(\textrm{SNR})}}$ as
the MG \cite{ref6}, where $R(\textrm{SNR})$ is the achievable sum-rate in the
$K$-user IC, and $\textrm{SNR}$ is defined as the total transmit power across
all transmitters. The frequency-selective channel realizations are called
channel uses or resources for convenience in this letter.
\subsection{Original Interference Alignment (OIA) in the $K$-User IC}
Let $N=(K-1)(K-2)-1$, $M=(n+1)^N+n^N$ ($n$ is a positive integer), and let
$(d^{[1]},d^{[2]},\dots,d^{[K]})=((n+1)^N,n^N,\dots,n^N)$ be the numbers of
streams allocated to the $K$ users respectively. Then
\begin{multline}\label{eq2}
\{r^{[1]},r^{[2]},\dots,r^{[K]}\}=\bigg\{\frac{(n+1)^N}{(n+1)^N+n^N},\\
\frac{n^N}{(n+1)^N+n^N},\dots,\frac{n^N}{(n+1)^N+n^N}\bigg\}
\end{multline}
are the achievable MGs of the $K$-user IC over $M$ channel uses. So, the total
achievable MG over $M$ channel uses is \cite{ref3}
\begin{equation}\label{eq3}
r_{OIA}(K,M)=\sum_{i=1}^K{r^{[i]}}=\frac{(n+1)^N+(K-1)n^N}{(n+1)^N+n^N}.
\end{equation}
\subsection{Beamforming Optimized Interference Alignment (BF-IA)}
An efficient IA scheme is proposed in \cite{ref4}, where the precoding matrices
are optimized. The streams allocated to the users are
\begin{multline}\label{eq4}
\{d^{[1]},d^{[2]},\dots,d^{[K]}\}=\\
\left\{ \binom{n^*+N+1}{N},\binom{n^*+N}{N},\dots,\binom{n^*+N}{N}\right\},
\end{multline}
and the dimension of the extended channel is
\begin{equation}\label{eq5}
M=d^{[1]}+d^{[2]}=\binom{n^*+N+1}{N}+\binom{n^*+N}{N},
\end{equation}
where $N=(K-1)(K-2)-1$ and $n^{*}$ is a nonnegative integer. The total
achievable MG is \cite{ref4}
\setlength{\arraycolsep}{0.0em}
\begin{eqnarray}
r_{BF}(K,M)=\frac{d^{[1]}+(K-1)d^{[2]}}{d^{[1]}+d^{[2]}}\label{eq6} ~~~~~~~~~~~~~~~~~\\
=\frac{(K-1)(n^*+1)+n^*+N+1}{2n^*+N+2}.\label{eq7}
\end{eqnarray}
\setlength{\arraycolsep}{5pt}
For example, when $K=4$, $N=(4-1)(4-2)-1=5$, a
solution to IA is feasible over the following dimensions of the extended
channel: $\mathcal{L}_4=\{7,27,77,182,\dots\}$. $\mathcal{L}_k$ is defined as
the set of all the feasible length of the extended channel over $k$ users when
$k\geq 3$. When $k <3$, orthogonal multiplexing is MG optimal. So, let
$\mathcal{L}_1=\mathcal{L}_2=\{1,2,3,\dots\}$.

When $M\to\infty$, we have
\begin{equation}
\lim_{M\to\infty}r_{BF}(K,M)=\lim_{n^*\to\infty}\frac{Kn^*+K+N}{2n^*+N+2}=\frac{K}{2}.
\end{equation}
Hence, allocating all the resources to all users simultaneously is MG optimal.

\begin{figure}[!tr]
\centering
\includegraphics[scale=.64]{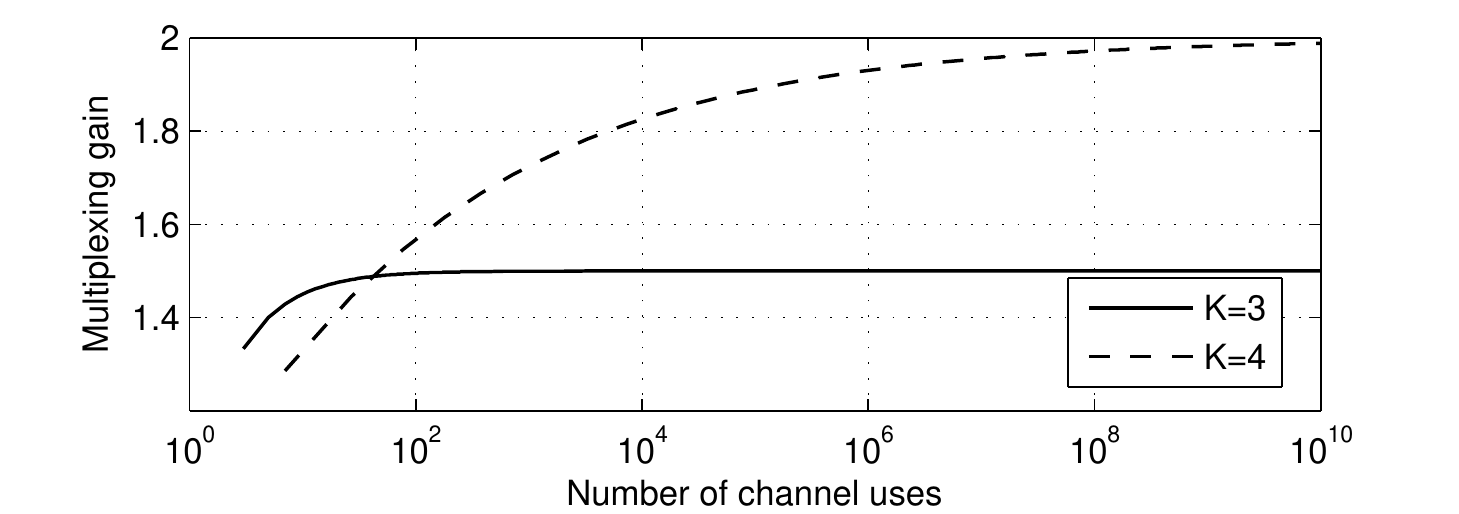}
\caption{The achievable MG of the BF-IA scheme when $K=3$ and $K=4$.}
\label{fig1}
\end{figure}

Using (\ref{eq6}), Fig.~\ref{fig1} illustrates the achievable MG when $K=3$ and
$K=4$. It can be seen that when $M\geq 77$ allocating all the resources to all
users simultaneously ($K=4$) can obtain more MG, and when $M<77$ allocating
them to partial users $(K=3)$ can obtain more MG. For example, when $M=7$,
$r_{BF}(4,7)\approx1.2857$ while $r_{BF}(3,7)\approx1.4286$.

For general values of $M$ and $K$, a natural question is how to allocate
resources among users can obtain more MG. In the following section a GIA scheme
is proposed based on the BF-IA scheme.
\section{Group Based Interference Alignment}
\begin{figure}[!tr]
\centering
\includegraphics[width=7.7cm,height=3.0cm]{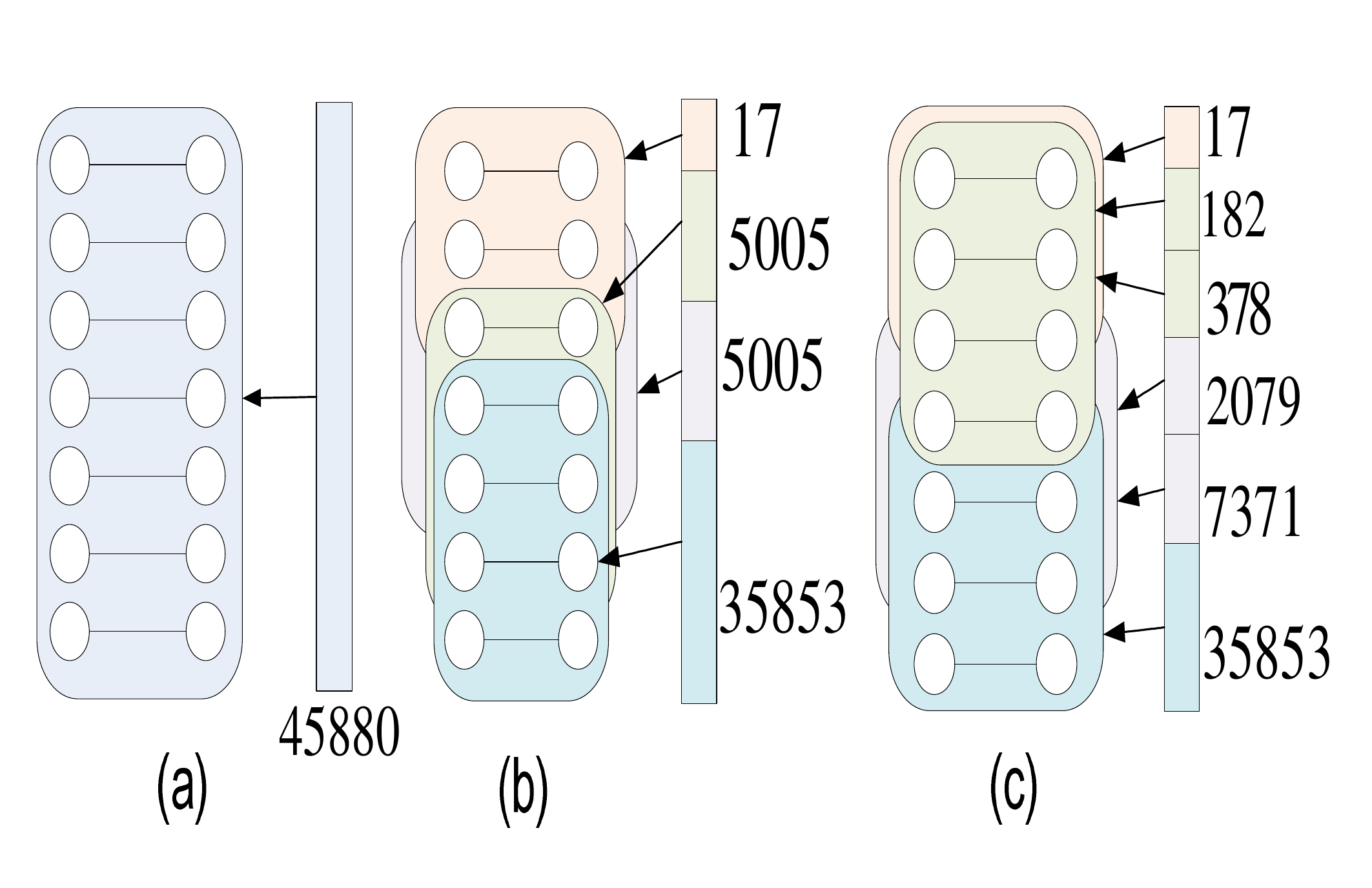}
\caption {Comparison of the BF-IA and the GIA schemes when $K=7$ and $M=45880$.
(a) The BF-IA scheme, which allocates all the resources to $7$ users
simultenously. (b) The GIA scheme (optimal algorithm), where $M$ is divided
into 4 parts ($m_1 = 35853$, $m_2 = m_3 = 5005$, and $m_4 = 17$). (c) The GIA
scheme (greedy algorithm), where $M$ is divided into 6 parts ($m_1 = 35853$,
$m_2 = 7371$, $m_3 = 2079$, $m_4 = 378$, $m_5 = 182$, and $m_6 = 17$).}
\label{fig2}
\end{figure}

Let $M$ be the total resources that can be used for IA in the $K$-user IC. Let $k\leq K$, $m \in \mathcal{L}_k$ and $m \leq M$, and let
\begin{equation}
v=\begin{cases}
\frac{m}{M}[r_{BF}(k,m)-1] & k\geq 3\\
0 & k< 3
\end{cases}
\end{equation}
be the relative MG that the BF-IA scheme obtained compared to orthogonal
multiplexing scheme (when $k<3$, orthogonal multiplexing is MG optimal). Define $e=\{k,m,v\}$ as a group pattern. For
example, in Fig.~\ref{fig2}a we have $e=\{7,45880,0.5405\}$ while in Fig.~\ref{fig2}b we have $e_1=\{4,35853,0.6759\}$,
$e_2=e_3=\{4,5005,0.0873\}$, and $e_4=\{3,17,0.0002\}$.

Let $\mathcal{S}_k^M = \{\{k,m_1,v_1\}, \{k,m_2,v_2\},\dots,\{k,m_w,v_w\}\}$ be
the set of all the group patterns over $k$ users exactly, where $m_j \leq M$, $1 \leq j
\leq w$. Define the relative MG obtained per dimension as
\begin{equation}
\rho_j =\begin{cases}
\frac{v_j}{m_j}= \frac{1}{M}[r_{BF}(k,m_j)-1] & k\geq 3\\
0 & k<3
\end{cases}
\end{equation}
which is the \emph{efficiency} of a group pattern. For example, in Fig.~\ref{fig2}a we have $\rho \approx 1.1782\times 10^{-5}$ while in
Fig.~\ref{fig2}b we have $\rho_1 \approx 1.8851\times 10^{-5}$, $\rho_2=\rho_3
\approx 1.7437\times 10^{-5}$, and $\rho_4 \approx 1.0257\times 10^{-5}$.

Let $\mathcal{E}_K^M = \bigcup_{j=1}^K \mathcal{S}_j^M$ be the set of all the
group patterns over any $k$ users when $k \leq K$. We denote the elements of
$e_i$ as $e_i.k$, $e_i.m$, and $e_i.v$ respectively. If there exist two group
patterns $e_i$ and $e_l$ with $e_i.m \leq e_l.m$ and $e_i.v \geq e_l.v$ in the set
$\mathcal{E}_K^M$, then it would always be better (or at least not worse) to
choose $e_i$. Hence, $e_l$ is removed from the set $\mathcal{E}_K^M$, and we
obtain $\mathcal{\overline{E}}_K^M = \{e_1,\dots,e_W \}$,
$W=|\mathcal{\overline{E}}_K^M|$. For example, if $e_i=\{3,7,6.5388\times
10^{-5}\}\in \mathcal{S}_3^{45880}$ and $e_l= \{4,7,4.3592\times 10^{-5}\}\in
\mathcal{S}_4^{45880}$, then $e_l$ is removed from the set $\mathcal{E}_K^M$.

The elements of $\mathcal{\overline{E}}_K^M$ are sorted by $\rho$ in
non-increasing order, and we obtain $\overleftarrow{\mathcal{E}}_K^M =
\{e_1,\dots,e_W \}$. For example, when $K=7$ and $M=45880$, we have
$\overleftarrow{\mathcal{E}}_K^M =\{\{4,35853,0.6759\}$, $\{4,27132,0.5069\}$,
$\{4,20196,0.3735\}$, $\{5,44200,0.8092\},\dots\}$
 where $\rho_1 \approx  1.8851\times 10^{-5}$, $\rho_2 \approx
1.8682\times 10^{-5}$, $\rho_3 \approx  1.8494\times 10^{-5}$, $\rho_4 \approx
1.8309\times 10^{-5}$, $\dots$, and $W=470$.

Given $K$ and $M$, we have obtained all the group patterns sorted by
efficiency, and the question becomes how to choose among them so as to obtain
more MG. It is modeled as an unbounded knapsack problem which is NP-hard
\cite{ref7}. The corresponding integer programming formulation is given as
follows.
\begin{align}
    \max~\sum_{j=1}^W x_j\cdot e_j.v~~~~~~~\,\; \label{eq8} \\
    \textrm{subject~to}~\sum_{j=1}^W x_j\cdot e_j.m \leq M \label{eq9}\\
    x_j \geq 0, \textrm{integer},j=1,\dots,W.\label{eq10}
\end{align}
We denote the solution values by $z_{o}$ (optimal algorithm) and $z_{g}$
(greedy algorithm), and denote the solution sets by $\mathcal{P}_{o}$ (optimal
algorithm) and $\mathcal{P}_{g}$ (greedy algorithm) respectively.

\subsection{Optimal Search Algorithm}
As a standard dynamic programming algorithm, Unbounded-DP \cite{ref7} is
adapted to evaluate our problem.
\\$\overline{~~~~~~~~~~~~~~~~~~~~~~~~~~~~~~~~~~~~~~~~~~~~~~~~~~~~~~~~~~~~~~~~~~~~~~~}$\\
$\textbf{Algorithm}~\textrm{optimal}$

for $m:=0$ to $M$ do

\quad $z(m):=0$, $r(m):=0$ ~~~~~~~~~~~~~\%~\emph{initialization}

for $j:=1$ to $W$ do

\quad for $m:=e_j.m$ to $M$ do~~~~~~~~~~~~\%~\emph{$e_j$ may be packed}

\quad\quad if $z(m-e_j.m) + e_j.v \geq z(m)$ then

\quad\quad\quad $z(m):=z(m-e_j.m)+e_j.v$

\quad\quad\quad $r(m):=j$

$\mathcal{P}:=\varnothing$, $\overline{m}:=M$

repeat~~~~~~~~~~~~~~~~~~~~\%~\emph{recover the optimal solution set}

\quad $r:=r(\overline{m})$

\quad $\mathcal{P}:= \mathcal{P} \bigcup \{e_r\}$

\quad $\overline{m}:=\overline{m}-e_r.m$

until $\overline{m} = 0$

$z_o:=z(M)$, $\mathcal{P}_{o}:= \mathcal{P}$
\\$\overline{~~~~~~~~~~~~~~~~~~~~~~~~~~~~~~~~~~~~~~~~~~~~~~~~~~~~~~~~~~~~~~~~~~~~~~~}$

Let $K = 7$, $M= 45880$ ($\in \mathcal{L}_7$). Fig.~\ref{fig2}b shows the
search results of the optimal algorithm. The total MG obtained is
 $r_G^{opt} = z_o +1 \approx1.8506$. Using (\ref{eq6}), the MG obtained by the
BF-IA scheme is about $1.5405$. So, about $20\%$ more MG is obtained by making
full use of $45880$ channel uses.

The computation complexity of this optimal algorithm is $O(MW)$, and it is a
pseudopolynomial algorithm \cite{ref7}.
\subsection{Greedy Search Algorithm}
When $M$ is large, the optimal algorithm will need prohibitive time to obtain
the solutions even using a powerful computer. So, a greedy algorithm is
proposed in the following:
\\$\overline{~~~~~~~~~~~~~~~~~~~~~~~~~~~~~~~~~~~~~~~~~~~~~~~~~~~~~~~~~~~~~~~~~~~~~~~}$\\
$\textbf{Algorithm}~\textrm{greedy}$

$m:=0$, $z:=0$, $\mathcal{P}:=\varnothing$

for $j:=1$ to $W$ do

\quad if $m+e_j.m \leq M $ then

\quad\quad $x_j:=\lfloor (M-m)/e_j.m \rfloor$

\quad\quad $m:= m + x_j\cdot e_j.m$

\quad\quad $z:= z + x_j\cdot e_j.v $

\quad\quad $\mathcal{P}:= \mathcal{P} \bigcup
\{\underbrace{e_j,\dots,e_j}_{x_j}\}$

$z_g=z$, $\mathcal{P}_g = \mathcal{P}$
\\$\overline{~~~~~~~~~~~~~~~~~~~~~~~~~~~~~~~~~~~~~~~~~~~~~~~~~~~~~~~~~~~~~~~~~~~~~~~}$

The computation complexity of the greedy algorithm is $O(W)$. However, as the
greedy algorithm does best every step which is locally optimal, global optimum
is not confirmed. According to \emph{Theorem} 8.5.1 in \cite{ref7}, the
relative performance guarantee ($z_g/z_o$) of the greedy algorithm is bounded
by $1/2$. Fig.~\ref{fig2}c shows the search results of the greedy algorithm
when $K=7$ and $M=45880$. In this example, the total MG obtained is
$r_G^{grd}=z_g + 1 \approx 1.8494$, and the relative performance guarantee is
about $99.86\%$.
\subsection{Discussions}
When each user is equipped with $T$ antennas, let $K^\prime = KT$, $N^\prime
=(K^\prime-1)(K^\prime-2)-1$, and
$M^\prime=\binom{n^*+N^\prime+1}{N^\prime}+\binom{n^*+N^\prime}{N^\prime}$.
Using \emph{Corollary} $1$ in \cite{ref4}, the achievable MG is
$r_{BF}(K^\prime,M^\prime)$. So, the GIA scheme and the search algorithms can
be readily extended to this scenario.

The stream allocations among users are non-uniform both in the BF-IA and the
GIA schemes. Hence, dynamic algorithm should be considered to balance
unfairness among users. There is a simple solution, that is choosing the user
served with more streams from all users periodically. Also, sophisticated
schemes can be designed to incorporate other factors. However, we leave it for
future work.
\section{Numerical Results}
\begin{figure}[!tr]
\centering
\includegraphics[scale=.60]{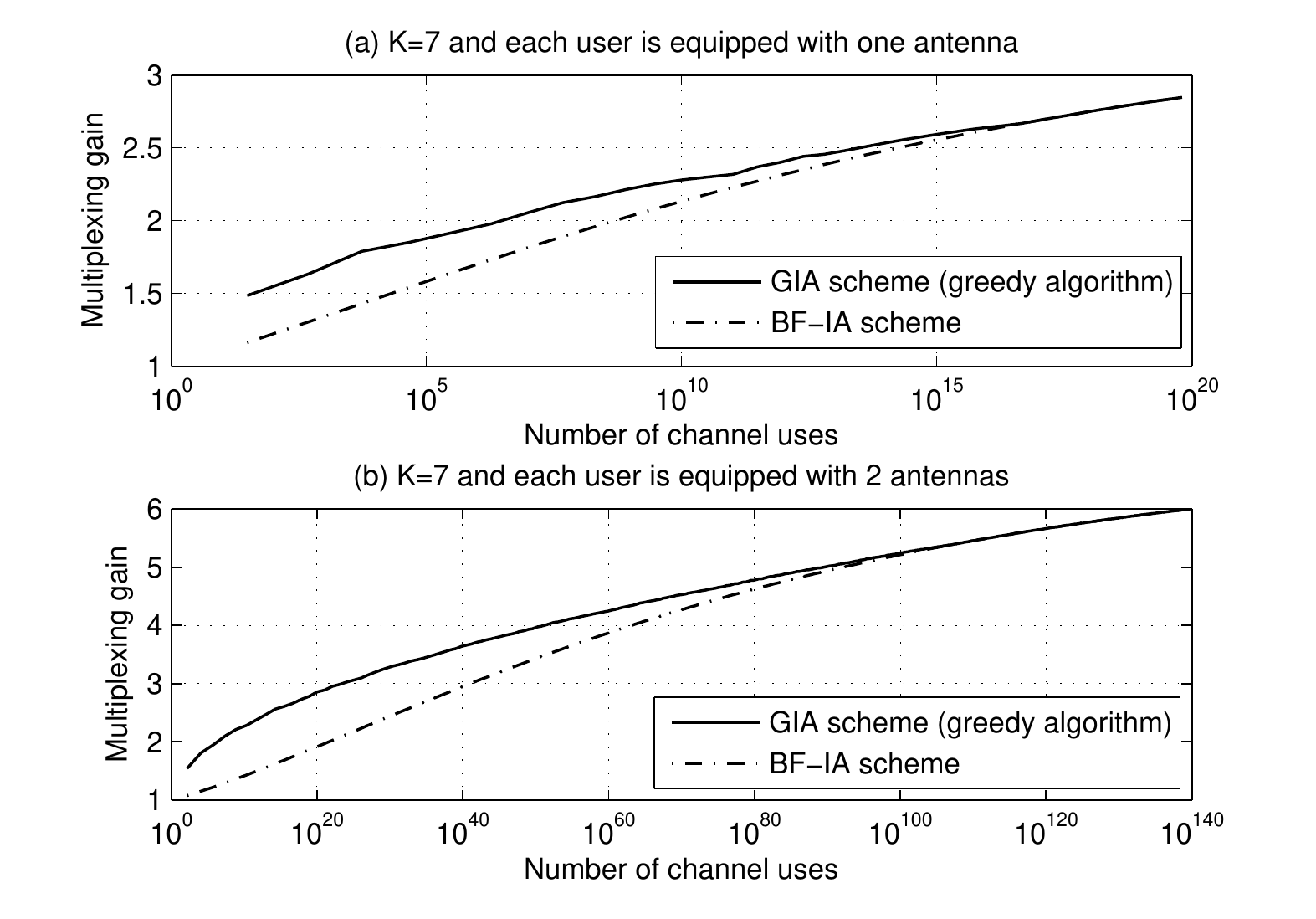}
\caption{Comparison of the GIA (greedy algorithm) and the BF-IA schemes in the
$K$-user IC. (a) $K=7$ and each user is equipped with one antenna. (b) $K=7$
and each user is equipped with 2 antennas, i.e., $K^\prime=KT=14$.}
\label{fig3}
\end{figure}

The comparison of the GIA and the BF-IA schemes in the $K$-user IC is presented
in Fig.~\ref{fig3}. We only compare the achievable MG when $m \in
\mathcal{L}_7$ in Fig.~\ref{fig3}a (or $m \in \mathcal{L}_{14}$ in
Fig.~\ref{fig3}b), as the BF-IA scheme has no solution when $m \notin
\mathcal{L}_7$ (or $m \notin \mathcal{L}_{14}$). And we also only compare the
greedy algorithm of the GIA scheme with the BF-IA scheme, as the optimal
algorithm is only feasible when $M$ is small.

Fig.~\ref{fig3} shows that the proposed GIA scheme obtains a higher MG when the
resources below a certain value (e.g., about $10^{17}$ in Fig. \ref{fig3}a).
When the resources are large, as discussed above, allocating all them to all
users simultaneously is MG optimal. The unsteadiness of the curves of the GIA
scheme is because the discrete nature of the problem. The GIA scheme is
preferred to be used when the resources are limited.

\section{Conclusion}
In this letter, under the same IA conditions as in \cite{ref4}, a group based
IA scheme is proposed. Analysis and numerical results show that the GIA scheme
obtains a higher MG in comparison with the BF-IA scheme when the resources are
limited.

\ifCLASSOPTIONcaptionsoff
  \newpage
\fi

\begin{thebibliography}{100}

\bibitem{ref1}
M.~Maddah-Ali, A.~Motahari, and A.~Khandani, ``Signaling over MIMO multi-base
systems: Combination of multi-access and broadcast schemes," \emph{in Proc.
IEEE Int. Symp.Inform.Theory}, 2006.

\bibitem{ref2}
M.~Maddah-Ali, A.~Motahari, and A.~Khandani, ``Communication over MIMO X
channel: Interference alignment, decomposition, and performance analysis,"
\emph{IEEE Trans. Inf. Theory}, vol. 54, no. 8, pp. 3457-3470, Aug. 2008.

\bibitem{ref3}
V.~R.~Cadambe and S.~A.~Jafar, ``Interference alignment and the
degrees of freedom for the K user interference channel," \emph{IEEE
Trans. Inf. Theory}, vol. 54, no. 8, pp. 3425-3441, Aug. 2008.

\bibitem{ref4}
S.~W.~Choi, S.A.~Jafar, and S.Y.~Chung, ``On the beamforming design
for efficient interference alignment," \emph{IEEE Commun. Lett.},
vol. 13, no. 11, pp. 847-849, Nov. 2009.

\bibitem{ref5}
C.~Suh, M.~Ho, and D.~Tse, ``Downlink interference alignment," in
arXiv:cs.IT/1003.3707v2, May. 2010.

\bibitem{ref6}
L.~Zheng and D.~N.~C.~Tse, ``Diversity and multiplexing: a
fundamental tradeoff in multiple-antenna channels," \emph{IEEE
Trans. Inf. Theory}, vol. 49. no. 5, pp. 1073-1096, May. 2003.

\bibitem{ref7}
H.~Kellerer, U.~Pferschy, and D.~Pisinger, \emph{Knapsack Problems}, Springer,
2004.
\end{thebibliography}
\end{document}